\documentclass[twocolumn]{article}
\usepackage{amssymb}
\usepackage{amsmath}

\newcommand{\tr}{\mathop{\rm tr}\nolimits}
\newcommand{\X}{\mathcal{X}}
\newcommand{\M}{\mathcal{M}}

\newcommand{\calO}{\mathcal{O}}

\newcommand{\x}{\mathbf{x}}

\newcommand{\bfu}{\mathbf{u}}
\newcommand{\vxi}{\vec{\xi}}

\newcommand{\R}{\mathbb{R}}

\title{Universal approach to gravitational thermal effects}
\author{Michael B. Mensky\\
{\small P.N.Lebedev Physical Institute, 53 Leninsky prosp., 117924 Moscow, Russia}\\
\centerline{\small and}\\
{\small Institut f\"ur Theoretische Physik,
Technische Universit\"at Berlin}\\
{\small 36 Hardenbergstr., D-10623 Berlin, Germany}}
\date{}

\begin{document}

\maketitle

\begin{abstract}
A universal scheme for describing gravitational thermal effects is developed as a generalization of Unruh effect. Quasi-Rindler (QR) coordinates are constructed in an arbitrary curved space-time in such a way that the imaginary QR time be periodical. The observer at rest in QR coordinates should experience a thermal effect. Application to de Sitter space-time is considered.
\end{abstract}

\section{Introduction}
\label{sec:Introduction}

It is well known that thermal effects may arise in gravitational fields. Hawking effect is well-known \cite{Hawking76}: an observer remaining at constant distance from a black hole should detect particles with thermal energy distribution issued by the black hole. Unruh considered \cite{Unruh76} the thermal effect which should be experienced by a linearly accelerated observer in Minkowski space-time: vacuum should appear for such an observer as a thermal bath with the temperature proportional to the observer's acceleration.

Nature of Unruh effect has been extensively explored (see for example \cite{UnruhEffectLit}) although ultimate clarity in this question is not yet achieved \cite{UnruhProbl}. Its mathematical aspects are simple and transparent. This is why this effect is often used as a model for exploring other gravitational thermal effects. Unruh effect is conveniently described by the so-called Rindler coordinate frame $(\eta, \xi)$ in the Minkowski plane $(x^0, x^3)$. Two other Minkowski coordinates $(x^1, x^2)$ are irrelevant.

Rindler coordinates are defined in such a way that the lines $\vec{\xi} = (x^1, x^2, \xi) = \mathrm{const}$ are trajectories of accelerated observers, while \emph{Rindler time} $\eta$ is proportional to the length measured along each of these trajectories, or the observer's proper time. Rindler coordinates cover a part of the plane $(x^0, x^3)$ called first Rindler wedge $R_{\mathrm{I}}$ and restricted by the future and past event horizons $x^3 = \pm x^0$ of the observer.

Unruh effect is essentially connected with periodicity of the so-called \emph{imaginary Rindler time} $\theta$ defined by the analytical continuation (Wick rotation) of Rindler time onto imaginary values: $\eta = i\theta$, $\theta\in\R$.

We shall develop a scheme (called \emph{Universal Rindler Scheme}) for constructing an analogous \emph{quasi-Rindler} (QR) \emph{coordinate frame} $(\eta, \vec{\xi})$ in an arbitrary curved space-time. The scheme is essentially based upon the formalism of Path Group \cite{PathGroup} although it will be formulated in simpler terms.

Just as in Minkowski space-time, imaginary QR time $\theta$ in a curved space-time (defined by $\eta = i\theta$, $\theta\in\R$) is periodical. The observers moving along trajectories $\vec{\xi} = \mathrm{const}$ (at rest in QR coordinates, or moving along the lines of QR time) should therefore experience thermal effects. If the space-time metric is static in respect of QR time $\eta$ (i.e. the metric does not depend of $\eta$ and the components $g^{0k}$, $k = 1, 2, 3$, of this metric are zero), then these observers should perceive vacuum\footnote{the global vacuum defined with the help of \emph{quasi-Cartesian} coordinates} as a thermal bath with a definite temperature. 

Universal Rindler Scheme will be applied in de Sitter space-time. The trajectories leading to thermal effects in this space-time will be constructed and corresponding temperatures found. The minimum temperature perceived by the inertial observers coincides with the `temperature of de Sitter space-time'.

In Sect.~\ref{sec:MinkowskiRindler}  Rindler coordinates in Minkowski space-time are introduced. In Sect.~\ref{sec:PathGroup} the formalism of Path Group is sketched as a background for the proposed scheme. In Sect.~\ref{sec:RindlerScheme} Universal Rindler Scheme is formulated. In Sect.~\ref{sec:TemperGreenFunct} the connection of periodicity of imaginary time with thermal effects is discussed. In Sect.~\ref{sec:DeSitter} Universal Rindler Scheme is applied to de Sitter space-time, and in Sect.~\ref{sec:Conclusion} concluding remarks are made.

\section{Rindler coordinates}
\label{sec:MinkowskiRindler}

Interconnection of Rindler $(\eta, \xi)$ and Minkowski $(x^0, x^3)$ coordinates is following:
\begin{equation}
x^0=\ell_0 e^{\xi/\ell_0}\sin\frac{\eta}{\ell_0}, \quad
{x^3}=\ell_0 e^{\xi/\ell_0}\cosh\frac{\eta}{\ell_0}.
\label{MinkRindl}
\end{equation}
Two other coordinates $x^1$, $x^2$ are the same both in Minkowski and Rindler frames. The whole set of Rindler coordinates may be denoted as $(\eta, x^1, x^2, \xi) = (\eta, \vec{\xi})$ with $\xi^1 = x^1$, $\xi^2 = x^2$, $\xi^3 = \xi$. 

The Minkowski metric 
\begin{equation}
ds^2=e^{2\xi/\ell_0}(-d\eta^2+d\xi^2)+(d{x^1})^2+(d{x^2})^2.
\label{MetricRindlerPlain}\end{equation}
is static in respect of Rindler time $\eta$. Rindler coordinates cover the region  $(x^0)^2 - (x^3)^2 < 0$, $x^3 > 0$ of Minkowski space-time (called Rindler wedge $R_{\mathrm{I}}$).

The hyperbolic trajectory $\xi = 0$ or
\begin{equation}
\gamma_0\, : \quad
x^0=\ell_0\sinh\frac{\eta}{\ell_0}, \quad
{x^3}=\ell_0\cosh\frac{\eta}{\ell_0}
\label{AcceleratedTrajectory}\end{equation}
(with $(x^1, x^2) = (\xi^1, \xi^2) = \mathrm{const}$) describes the motion with acceleration $w = c^2/\ell_0$. Rindler time $\eta$ is a measure of length (proper time) along $\gamma_0$.

Unruh predicted \cite{Unruh76} that an observer moving along the trajectory $\gamma_0$ should detect around him a thermal bath with temperature $kT = \hbar c/{2\pi\ell_0}$ instead of vacuum. From mathematical point of view, the reason is periodicity of Rindler time $\eta$ with an imaginary period: $(\xi, \eta)$ and $(\xi, \eta + i\, 2\pi\ell_0)$ correspond to the same pair of Minkowski coordinates $(x^0,{x^3})$.

Another formulation of the same is periodicity, with a real period, of \emph{imaginary Rindler time}. Indeed, after Wick rotation (transition to imaginary time) $x^0 = i x^4$, $x^4\in\mathbb{R}$, or, in Rindler coordinates, $\eta = i\theta$, $\theta\in\mathbb{R}$, the connection between Minkowski and Rindler coordinates becomes
\begin{equation}
x^4=\ell_0 e^{\xi/\ell_0}\sin\frac{\theta}{\ell_0}, \quad
{x^3}=\ell_0 e^{\xi/\ell_0}\cos\frac{\theta}{\ell_0}.
\label{MinkImagRindl}
\end{equation}
Imaginary Rindler time $\theta$ is a polar angle in the plane $(x^4, x^3)$ and is periodical with real period ${2\pi}\ell_0$. 

This period determines temperature. To calculate it, one has to take into account that Rindler time $\eta$ is the \emph{proper time} of the observer moving along $\gamma_0$. After Wick rotation the hyperbola $\gamma_0$ is converted into a circumference of the length $\beta = {2\pi}\ell_0$. This length is equal to inverse temperature, $kT=\hbar c/\beta$.

If we consider a hyperbola $\xi = \mathrm{const}$ with arbitrary $\xi$ as an observer's trajectory (corresponding to the acceleration $w = c^2/\ell$ where $\ell = \ell_0 e^{\xi/\ell_0}$), then the length along this trajectory (i.e. the observer's proper time) is measured, according to Eq.~(\ref{MetricRindlerPlain}), by the parameter $e^{\xi/\ell_0}\eta$. After Wick rotation $\eta = i\theta$, $\theta\in\mathbb{R}$, the hyperbola $\xi = \mathrm{const}$ converts into a circumference of the length $\beta(\xi) = 2\pi \ell = 2\pi \ell_0\, e^{\xi/\ell_0}$. The temperature for the observer moving along the trajectory $\xi = \mathrm{const}$ will be therefore $kT(\xi)=\hbar c/\beta(\xi)$.

Universal Rindler Scheme to be proposed in Sect.~\ref{sec:RindlerScheme} allows one to construct Rindler-type coordinates in an arbitrary curved space-time. The most important feature of the resulting quasi-Rindler (QR) coordinates is periodicity of imaginary QR time providing thermal effects. 

\section{Path Group}
\label{sec:PathGroup}

The construction of a Rindler-type coordinate frame (Sect.~\ref{sec:RindlerScheme}) will be formulated in terms of geodesic lines specified by Minkowskian `guiding' four-vectors. However this is actually based on the formalism of Path Group \cite{PathGroup}. Let us sketch the concepts of Path Group which are behind the construction. This section may be skipped without detriment to understanding.

The elements of Path Group $P$ are classes of continuous curves in Minkowski space $\M$. The classes are composed in a special way to form a group. Each path $p\in P$ is presented by any curve from the corresponding class. Paths may be multiplied so that the curves extend each other, Path Group generalizing the group of translations. Unification of the group $P$ with the group of Lorentz transformations gives Generalized Poincar\'{e} Group $Q$.

Path Group and Generalized Poincar\'{e} Group may be applied for exploring (particles in) a curved (pseudo-Riemannian) space-time $\X$. Minkowski space $\M$ (in which the paths are defined) is then a model of tangent space to an arbitrary point $x\in\X$. This allows to naturally define the action of $P$ and $Q$ in the principal fiber bundle over $\X$. It is very important that Wick rotation of $P$ and of $Q$ (i.e. the group of paths in Euclidean 4-space and its unification with the group of 4-rotations) act in the (fiber bundle to) Euclidean slice of the space-time. 

The paths $p\in P$ (essentially, curves in $\M$) may serve as natural images of the curves in $\X$ provided that the starting point of the curve in $\X$ and a local frame $n$ in the starting point (four tangent vectors $n = \{ n_\alpha | \alpha = 0, 1, 2, 3\}$ in $x$) are given. The straight paths are mapped in this way onto geodesic lines. 

The straight paths in Minkowski space being in one-to-one correspondence with the four-vectors, each four-vector determines a geodesic in $\X$, given a point in $\X$ and a local frame in this point. More concretely, let a four-vector $a$, a starting point $\mathcal{O}$ of the geodesic and a local frame $n$ in in this point are given. Then the the geodesic should have tangent vector $a^\alpha n_\alpha$ in the point $\mathcal{O}$ and the length equal to the length of $a$. The final point $x\in\X$ of the geodesic is thus unambiguously determined by the four-vector $a$. The components of $a$ may be taken to be coordinates of  $x$. This will be used in Sect.~\ref{sec:RindlerScheme}.

Correspondingly, geodesics in Euclidean slice of the space-time $\X$ may be presented by Wick rotated ($a^0 = ia^4$, $a^4\in\mathbb{R}$) Minkowskian four-vectors, i.e. by Eucliden 4-vectors $(a^4, \mathbf{a})$. It is important for the constructions of Sect.~\ref{sec:RindlerScheme} that Wick rotation of (non-compact) Lorentz group is a (compact) group of 4-rotations: non-periodic hyperbolic angles are converted into periodic trigonometric ones. QR time will be defined as a hyperbolic angle, hence periodicity of imaginary QR time and thermal effects.

\section{Universal Rindler Scheme}
\label{sec:RindlerScheme}

Let us now describe the scheme of constructing Rindler-type coordinates which will be called \emph{quasi-Rindler} (QR) coordinates. The scheme is universal since it may be applied in an arbitrary (pseudo-Riemannian) space-time. Call it \emph{Universal Rindler Scheme}. 

We shall present the scheme assuming that the space-time geometry (metric) is known. If it is possible to construct QR coordinates in such a way that the metric is static in respect of QR time, the observer at rest in this frame should observe a thermal bath of definite temperature (otherwise the effect is more complicated). Alternatively, geometry may be specified after QR coordinates are introduced. In this case the metric may be chosen to be static.

Rindler coordinates are defined by Eq.~(\ref{MinkRindl}) in the plane $(x^0, x^3)$, while $x^1$, $x^2$ are fixed. One may say that a two-dimensional surface of points $(0, x^1, x^2, 0)$, $(x^1, x^2) \in \R^2$,  is determined and then from each point of this surface an orthogonal two-dimensional surface is built, with Rindler coordinates in it.

Universal Rindler Scheme will have an analogous structure. First we shall define a two-dimensional surface $\Sigma_{12}$, the points of which will be denoted $\mathcal{O}(x^1, x^2)$. Then from each of these points as an origin the two-dimensional `quasi-Rindler' surface $\Sigma_{03}$ will be built with QR coordinates $(\eta, \xi)$ in it.

For convenience we shall introduce 3-dimensional notation $\vxi$ for some of these parameters: $(\xi^1, \xi^2, \xi^3) = (x^1, x^2, \xi)$. The set of four parameters $(\eta, x^1, x^2, \xi) = (\eta, \vxi)$ will be called 4-dimensional QR coordinates. Besides this, we shall define in the surface $\Sigma_{03}$ the coordinates $(x^0, x^3)$ similar to Cartesian ones. Together with the coordinates $(x^1, x^2)$ of the origin this results in the coordinates $(x^0, x^1, x^2, x^3) = (x^0, \x)$ for the whole space-time called quasi-Cartesian.

In order to give the concrete definitions of these coordinate frames, we shall issue geodesic lines from each of the origins $\mathcal{O}(x^1, x^2)\in\Sigma_{12}$. To characterize the geodesics, we choose a local frame $n(x^1, x^2)$ in each point of $\Sigma_{12}$. Each of the local frames $n$ consists of four tangent vectors $n_\alpha$, $\alpha = 0, 1, 2, 3$, forming an orthonormal basis of the corresponding tangent space. Vectors $n_1$ and $n_2$ should be tangent to the surface $\Sigma_{12}$, while $n_0$ and $n_3$ orthogonal to it. The surface $\Sigma_{03}$ will turn out to be tangent to $n_0$ and $n_3$.

Given an origin $\calO$ and the local frame $n = \{ n_\alpha \}$ in it, the geodesic from $\calO$ in the direction of a (Minkowskian) four-vector $a$ is defined as follows. It starts in $\calO$, has in this point the tangent vector $a^\alpha n_\alpha$ and the length equal to the length of $a$. The vector $a$ is called \emph{guiding vector} of the geodesic.

Now we are ready to define both coordinate frames. Take an arbitrary point in $\Sigma_{12}$, say $\mathcal{O} = \mathcal{O}(x^1, x^2)$, together with the local frame in this point, $n = n(x^1, x^2)$, and do the following. 

Issue from $\mathcal{O} = \mathcal{O}(x^1, x^2)$ a geodesic line having the guiding vector
$
	a(x^0, x^3) = (x^0, 0, 0, x^3)
$
and associate the final point $x\in\X$ of this geodesic with the set of coordinates $(x^0, x^1, x^2, x^3)$ = $(x^0, \x)$. Call them \emph{quasi-Cartesian coordinates}. We shall assume that the coordinate frame obtained in this way, covers the whole space-time. 

We can now introduce \emph{quasi-Rindler coordinates} $(\eta, \vxi)$ with the same realtions as in Minkowski space, i.e. $\xi^1 = x^1$, $\xi^2 = x^2$, $\xi^3 = \xi$ and 
\begin{equation}
x^0 = \ell_0 \, e^{\xi/\ell_0}\sinh\frac{\eta}{\ell_0}, \quad
x^3 = \ell_0 \, e^{\xi/\ell_0}\cosh\frac{\eta}{\ell_0}. 
\label{quasiRindlerCartesian}\end{equation}
The QR coordinates always cover only a part of the space-time restricted by event horizons of the observers moving along the trajectories $\vxi = \mathrm{const}$. If it is possible, we shall try to choose the surface $\Sigma_{12}$ of origins in such a way that the space-time metric is static in respect of QR time $\eta$. 

An important remark has to be made about both coordinate frames. Depending on geometry of space-time $\X$, it may happen that two different sets of coordinates (say, $(x^0, \x)$ and $(x'^0, \x')$ or $(\eta, \vxi)$ and $(\eta', \vxi')$) correspond in fact to the same point of $\X$. This happens if the geodesic lines determined by these coordinate sets lead to the same point. It is essential however that for some vicinity $\mathcal{V}(x)$ of each point $x\in\X$ there exist a region of the space $\R^4$ of coordinate values which is in one-to-one correspondence with $\mathcal{V}(x)$. Then both sets of parameters may be used as multi-valued coordinates. This is enough for our aim.

As a consequence of the relation (\ref{quasiRindlerCartesian}), QR time $\eta$ is periodic with imaginary period: the sets $(\eta, \vec{\xi})$ and $(\eta + 2\pi i\ell_0, \vec{\xi})$ correspond to the same set of quasi-Cartesian coordinates $(x^0, \x)$ and therefore to the same point in space-time $\X$. Wording in another way, imaginary QR time $\theta$ (defined by $\eta = i\theta$, $\theta\in\R$) is periodical with real period $2\pi\ell_0$. The group-theoretical explanation of the periodicity is that a hyperbolic angle $\eta$ characterizing boosts is converted, under Wick ratation, into a trigonometric angle. 

Just as in Minkowski space-time with Rindler coordinates, periodicity of imaginary QR time $\theta$ leads to a thermal effect of the type of Unruh effect provided that the metric is static in respect of QR time $\eta$ (see Sect.~\ref{sec:TemperGreenFunct} for details). This means that vacuum\footnote{the vacuum defined by quasi-Cartesian coordinates $(x^0, \x)$ i.e. one corresponding to the Green function regular in Euclidean slice of the space-time as it is defined by $(x^4, \x)$} appears as a thermal bath for the observer moving along the trajectory $\vxi = \mathrm{const}$.

In order to find the temperature corresponding to the given trajectory $\vxi = \vxi_0$, one has to characterize the periodicity of QR time on this trajectory in terms of proper time. For this end one has to apply Wick rotation $\eta = i \theta$ on the trajectory and find the length $\beta$ of the resulting closed curve $\vxi = \vxi_0$ in the Euclidean slice of space-time. Then $kT = \hbar c/\beta$.

\section{Imaginary time, vacua and temperature}
\label{sec:TemperGreenFunct}

Let us discuss why periodicity of imaginary time leads to a thermal effect. We shall show that if the causal propagator of a particle is periodic in imaginary time, it has the form of temperature Green function \cite{Roepstorff94}. This means that the corresponding observer will see a thermal bath. It will be assummed that the space-time is curved but the metric is static in respect of the time coordinate $\eta$.

Consider a (scalar) particle propagating in this space-time. The metric being static, the coordinate $\eta$ is separable in the field equations from the space-like coordinates $\vxi$ and the positive-frequency solutions of the field equations may be taken in the form
$$
\psi_k(\eta,\vxi) = e^{-iE_k \eta}\varphi_k(\vxi)
$$
with $E_k > 0$. The field operator 
$
\Phi(\eta,\vxi) = \sum_k
\left[\psi_k(\eta,\vxi) a_k^\dag + \psi_k^\ast(\eta,\vxi) a_k\right]
$
(where $a_k$ are annihilation operators) gives for the transition amplitude
\begin{align*}
&\Delta(\eta,\vxi | \eta', \vxi')
= \langle 0_\mathrm{QR} |
\Phi(\eta,\vxi) \Phi(\eta',\vxi')
| 0_\mathrm{QR} \rangle \\
&= \Delta(\eta-\eta'|\vxi, \vxi')
= \sum_k e^{iE_k(\eta-\eta')}\, \varphi_k(\vxi, \vxi')
\end{align*}
where $| 0_\mathrm{QR} \rangle$ is the vacuum annihilated by $a_k$ and $\varphi_k(\vxi, \vxi') = \varphi_k^\ast(\vxi)\varphi_k(\vxi')$. Causal propagator
\begin{equation}
\Delta^c(\eta|\vxi, \vxi')
= \langle 0_\mathrm{QR} |
T[\Phi(\eta,\vxi) \Phi(0,\vxi')]
| 0_\mathrm{QR} \rangle
\label{CausalProp}
\end{equation}
is equal to
\begin{align*}
&\Delta^c(\eta|\vxi, \vxi')
=  \vartheta(\eta) \Delta(\eta|\vxi, \vxi')
+ \vartheta(-\eta) \Delta(-\eta|\vxi', \vxi) \nonumber \\
&= \sum_k \left[
\vartheta(\eta) e^{iE_k \eta}\, \varphi_k(\vxi, \vxi')
+ \vartheta(-\eta) e^{-iE_k \eta}\, \varphi_k(\vxi', \vxi)
\right].
\end{align*}

Perform now Wick rotation (analytical continuation to imaginary time) $\eta = i\theta$, $\theta\in\R$. In the course of this transition the semi-axis $\eta>0$ goes over into $\eta = i\theta$ with $\theta>0$, while $\eta<0$ goes over into $\eta = i\theta$ with $\theta<0$. Therefore the Wick-rotated causal propagator (called Schwinger function) is\footnote{It is defined in Euclidean slice of the space-time which is a 4-dimensional Riemannian space with signature $(++++)$.}
\begin{align}
	&S(\theta|\vxi, \vxi') = \Delta^c(i\theta|\vxi, \vxi')\nonumber\\
	&= \vartheta(\theta) \Delta(i\theta|\vxi, \vxi')
	+ \vartheta(-\theta) \Delta(-i\theta|\vxi', \vxi) \label{SchwingerFunct} \\
	&= \sum_k \left[
	\vartheta(\theta) e^{-E_k \theta}\, \varphi_k(\vxi, \vxi')
	+ \vartheta(-\theta) e^{E_k \theta}\, \varphi_k(\vxi', \vxi)
	\right] .   \nonumber
\end{align}
Energies $E_k$ being positive, Schwinger function vanishes at infinity of imaginary time:
\begin{equation}
	\lim_{\theta\to\pm\infty}S(\theta|\vxi, \vxi') = 0.
\label{SchwingerVanish}
\end{equation}

Now we shall introduce periodicity of Schwinger function in imaginary time $\theta$. This will convert causal propagator into temperature Green function.

Schwinger function is a Green function, i.e. it satisfies (Euclidean form of) the field equation with delta-function in the right-hand side. The Green function (\ref{SchwingerFunct}) is singled out by condition (\ref{SchwingerVanish}). The Green function with period $\beta$ is obtained by summing: 
\begin{align}
&S_\beta(\theta|\vxi, \vxi')
= \sum_{n=-\infty}^{\infty} S(\theta+n\beta|\vxi, \vxi')
\nonumber \\
&= \sum_k
\frac{e^{-E_k\theta}\varphi_k(\vxi,\vxi') + e^{-E_k\beta}e^{E_k\theta}\varphi_k(\vxi',\vxi)}{1 - e^{-E_k\beta}}.
\label{PeriodSum}
\end{align}
Its back Wick rotation $\theta = -i\eta$ results now in
\begin{align}
&W_\beta(\eta|\vxi, \vxi')
= S_\beta(-i\eta|\vxi, \vxi')  \nonumber\\
&= \sum_k
\frac{e^{iE_k \eta}\varphi_k(\vxi,\vxi')
+ e^{-E_k\beta}e^{-iE_k \eta}\varphi_k(\vxi',\vxi)}{1 - e^{-E_k\beta}}.
\label{TemperatureGreenFunc}\end{align}
This function may be shown to coincide with the temperature Green function
$
\langle
\Phi(\eta,\vxi) \Phi(0,\vxi')
\rangle_\beta
$
defined by the thermal average
$
\langle A \rangle_\beta
= {\tr (e^{-\beta H} A)}/{\tr (e^{-\beta H})}
$
and the Hamiltonian $H = \sum_k E_k \, a_k^\dag a_k$. It is essential for interpretation (\ref{TemperatureGreenFunc}) as a thermal function that its arguments are close to the trajectory along which the parameter $\eta$ measures length (proper time of the corresponding observer).

Thus periodicity of Green function in imaginary time results in thermal character of the propagation. Yet we have to clearly distinguish between two different reasons for periodicity of Green function.

If imaginary time $\theta$ has topology of real axis $\R$, periodicity with any period $\beta$ may be imposed by the summation (\ref{PeriodSum}). Resulting temperature Green function describes propagating the particle in a real (consisting of real particles) thermal bath with temperature $kT = \hbar c/\beta$.

If imaginary time is periodic (has topology of circumference $\mathbb{S}^1$) because of geometrical reasons, then the Schwinger function is necessarily periodic, and the period $\beta$ is fixed. Just this happens with imaginary Rindler and quasi-Rindler (QR) time. The reason of thermal character of Green function is then different. It is that the time parameter $\eta$ (corresponding to the periodic imaginary time $\theta$) is not one in respect of which the vacuum is defined.

Let us explain this for Minkowski space-time with Rindler time and a curved space-time with QR time.

When it is claimed that ``vacuum appears for the accelerated observer as a thermal bath'', the Minkowski vacuum is meant. As it follows from the above arguments (see also for example \cite{Roepstorff94}), the causal propagator defined with the help of this vacuum corresponds to the Schwinger function decreasing at infinity of imaginary Minkowski time $x^4$. Vice versa, the Green function decreasing at infinity of $x^4$ is the analytical continuation of the causal propagator defined by the Minkowski vacuum. The condition for Green function to decrease at infinity of $x^4$ is a definition of the Minkowski vacuum. 

If however this Schwinger function is expressed in terms of parameter $\theta$ (imaginary Rindler time) with the help of Eq.~(\ref{MinkImagRindl}), it turns out to be periodical simply because $\theta$ is a polar angle in the plane $(x^4, x^3)$. The function obtained by back Wick rotation $\theta = -i \eta$ is nothing else than the usual (Minkowskian) causal propagator. However, being expressed through the Rindler time, it possesses thermal properties.

The same situation arises in Universal Rindler Scheme of Sect.~\ref{sec:RindlerScheme}. Two coordinate frames, quasi-Cartesian $(x^0, \x)$ and quasi-Rindler $(\eta, \vxi)$, are related by Eq.~(\ref{MinkImagRindl}). It is assumed that the vacuum $| 0 \rangle$ (which is actually around the observer) corresponds to the Schwinger function decreasing at infinity of $x^4$ (if this parameter has topology of $\R$) or, in a more general case, regular in Euclidean slice of the space-time as it is defined by $(x^4, \x)$.\footnote{The metric is not necessarily static in respect of $x^0$.} This function is periodical in $\theta$, therefore the causal propagator defined with the vacuum $| 0 \rangle$ possesses thermal properties for the observer at the trajectory $\vxi = \mathrm{const}$. The vacuum $| 0 \rangle$ of the quasi-Cartesian particles looks like a thermal bath of the QR particles (presented by $a_k$, $a_k^\dag$). 

The question may arise why the vacuum corresponding to quasi-Cartesian coordinates is preferred in this argument. The answer is that these coordinates cover the whole space-time (at least we assume the geometry to possess this property). Regularity of the Schwinger function in $(x^4, \x)$ means its regularity in the whole Euclidean slice of the space-time. The vacuum is `global' in the sense that the corresponding Schwinger function has right analytical properties in the whole (complexified) space-time. There are evidences that such global vacua are stable \cite{MBMvacuum}. 

On the contrary, quasi-Rindler coordinates cover only the quasi-Rindler wedge. The vacuum which could be connected with QR time (QR vacuum denoted above $| 0_\mathrm{QR} \rangle$, for which Schwinger function vanishes at infinity of imaginary QR time), would be unstable. It could exist around the observer on trajectory $\vxi = \mathrm{const}$ only if this observer, together with his laboratory, be isolated from the environment by opaque (for the particles in question) walls. 

One more question is whether any thermal effect should take place if the space-time metric is not static in respect of QR time $\eta$. First of all, it is evident that the effect will only slightly change if the metric is close to one satisfying this condition (is obtained by its small deformation). If the metric is far from being static, then the propagator of a particle in this space-time does not have the form of temperature Green function. Nevertheless, as a consequence of periodicity of QR time, the expression for this propagator includes the specific factor $e^{-\beta H}$ of Boltzman form, although in a more complicated way than in the thermal Green function. Therefore, even in this case a sort of a thermal effect will arise for an observer on trajectory $\vxi = \mathrm{const}$.

\section{De Sitter space-time}
\label{sec:DeSitter}

Let us apply Universal Rindler Scheme to construct quasi-Rindler (QR) coordinates in de Sitter space-time. The latter will be described as a hyperboloid
$$
-(u^0)^2+{\bf u}^2+(u^5)^2 = R^2
$$
in 5-dimensional pseudo-Euclidean space with the coordinates $(u^0, \bfu, u^5)$, $\bfu = (u^1, u^2, u^3)$, and metric
$$
ds^2 = -(du^0)^2 + d\bfu^2 + (du^5)^2.
$$

Geodesic lines in de Sitter space-time are `large circles' on the hyperboloid: intersections of it with hyperplanes passing through the hyperboloid's center. Trajectories of inertial observers are time-like geodesics. They will be obtained below among those providing thermal effects. One of them has the form  
\begin{equation}
u^1=u^2=u^5=0, \quad (u^0)^2-(u^3)^2=-R^2.
\label{TrajDeSitInertObserv}\end{equation}

We need no quasi-Cartesian coordinates in de Sitter space-time because the coordinates $(u^0, \bfu, u^5)$ in 5-dimensional space define the `global' vacuum: after Wick rotation $u^0 = iu^4$ the causal propagator should be regular on the 4-sphere $(u^4)^2+{\bf u}^2+(u^5)^2 = R^2$ which is an Eucledean slice of de Sitter space-time.

QR coordinates are built starting from the set of origins $\calO(\xi^1, \xi^2)$ forming a sphere. Consider first one of them, $\calO = \calO(0, 0)$, presented with the point $(u^0, \bfu, u^5) = (0, \vec{0}, R)$ on the hyperboloid. The local frame in $\calO$ consists of the tangent vectors to the coordinate lines $u^0$, $u^1$, $u^2$, $u^3$.

To begin constructing QR coordinates, define a geodesic starting in $\calO$ and having the 4-vector
$$
a=(a^0,a^1,a^2,a^3)=(0,0,0,\ell)
$$
as a guiding vector. This geodesic has the form
$$
u(\varphi)=(u^0, u^1, u^2, u^3, u^5)
=R(0, 0, 0, \sin\varphi, \cos\varphi).
$$
If length of the guiding vector (and of the geodesic) is taken to be $\ell=\pi R/2$, the final point of the geodesic $u(\pi/2)=(0, 0, 0, R, 0)$ lies on the trajectory (\ref{TrajDeSitInertObserv}) of the inertial observer. In case of arbitrary length $\ell$ the final point of the geodesic is $u(\ell/R)$.

In this way we have only a single point of the trajectory of observer (corresponding to zero QR time $\eta = 0$). To obtain all points of the trajectory, take a `boosted' guiding vector
$$
\lambda^{\eta} a = \ell(\sinh\eta, 0, 0, \cosh\eta).
$$
The geodesic line from $\calO$ in the direction of $\lambda^{\eta} a$ is
$$
u(\varphi) 
=R(\sinh\eta\sin\varphi, 0, 0, \cosh\eta\sin\varphi, \cos\varphi).
$$
The final point of this geodesic $u(\eta,\ell) = u(\ell/R)$, or 
\begin{align}
&u(\eta,\ell) 
\nonumber \\
&=R(\sinh\eta\sin\frac{\ell}{R}, 0, 0, \cosh\eta\sin\frac{\ell}{R}, \cos\frac{\ell}{R})
\label{TrajObserverDeSitter}\end{align}
is a point, corresponding to QR time $\eta$, on the trajectory $\ell = \mathrm{const}$.

Various values of $\eta$ and $\ell$ give us various points of the QR surface.\footnote{More precisely, they cover only a part of the QR surface $\Sigma_{03}$ analogous to the Rindler wedge $R_{\mathrm{I}}$ in Minkowski space.} The pair of parameters $(\eta, \ell)$, $\eta\in\R$, $\ell\in\R_+$ may be coordinates on this surface. More conventional QR coordinates $(\eta, \xi)\in\R^2$ include the logarithmic coordinate $\xi$ defined by $\ell=\ell_0e^{\xi}$. It is straightforward to show that the metric in the QR surface is static in respect of QR time $\eta$.

To expand QR coordinates onto a 4-dimensional region of de Sitter space-time, we have to build up an analogous QR surface starting from a generic `origin' $\calO(\xi^1, \xi^2)\in\Sigma_{12}$. It is obtained from $\calO(0, 0)$ (the point $(0, \vec{0}, R)$ of the hyperboloid) by rotating 3-vector $(u^1, u^2, u^5) = (0, 0, R)$ and taking the components $(u'^1, u'^2)$ of the rotated 3-vector as $(\xi^1, \xi^2)$. The local frame is rotated correspondingly.

After building QR surface with coordinates $(\eta, \xi)$ but starting from $\calO(\xi^1, \xi^2)$ we may ascribe coordinates $(\eta, \xi^1, \xi^2, \xi)$ to its points. The notation $\xi^3 = \xi$ gives 4-dimensional QR coordinates $(\eta, \vxi)$.

The observer moving along the trajectory $\vxi = \mathrm{const}$ should experience a thermal effect. The trajectory is a line (\ref{TrajObserverDeSitter}) on the QR surface (with parameter $\ell$ fixed). The temperature depends on $\ell$. To show this and find the temperature, perform Wick rotation $\eta = i \theta$, $u^0 = iu^4$ to imaginary QR time $\theta$. Then trajectory (\ref{TrajObserverDeSitter}) is converted into the curve
\begin{align}
&u_{\mathrm{Eucl}}(\theta,\ell) = (u^4, u^1, u^2, u^3, u^5)  \nonumber\\
&= R(\sin\theta\sin\frac{\ell}{R}, 0, 0,
 \cos\theta\sin\frac{\ell}{R}, \cos\frac{\ell}{R})
\label{ImaginaryTrajObserverDeSitter}\end{align}
in  Euclidean 5-dimensional space with metric
$$
ds^2 = (du^4)^2 + d\bfu^2 + (du^5)^2.
$$
The curve $\ell = \mathrm{const}$ is a circumference of the length (inverse temperature)
\begin{equation}
\beta(\ell) = \frac{1}{k\, T(\ell)} = 2\pi R \sin\frac{\ell}{R}.
\label{TemperatureDeSitter}\end{equation}
If $\ell=\pi R/2$, then Eq.~(\ref{TrajObserverDeSitter}) presents trajectory (\ref{TrajDeSitInertObserv}) of an inertial observer. The temperature detected by him $kT = \hbar c/2\pi R$ is usually called temperature of de Sitter space-time. Generic $\ell$ corresponds to an accelerated observer and higher temperature. If $\ell\to 0$, the temperature increases as $kT(\ell) \approx {\hbar c}/{2\pi\ell}$.

\section{Conclusion}
\label{sec:Conclusion}

Summing up, we have shown how one can, in an arbitrary space-time, construct such a coordinate frame which is analogous to Rindler coordinate frame in Minkowski space. Owing to a special feature of this frame (periodicity of the Wick-rotated parameter of time), the observer moving along any of its time coordinate lines should observe a sort of thermal effect.

If the space-time metric is static in respect of the time parameter of this coordinate frame, then the effect consists in that the vacuum (defined by the quasi-Cartesian coordinates) appears to this observer as a thermal bath of a certain temperature. If the metric is not static, then the effect will have some features which are specific for thermal effects because the propagator of a particle along the observer's trajectory includes a factor of Boltzman type.

Application of the results to de Sitter space-time gave all trajectories on which definite temperatures should be observed. In general case they are higher than the temperature $kT = \hbar c/2\pi R$ detected by inertial observers.

It can be shown that Hawking effect may also be derived in the framework of this approach.

\quad

\textbf{\Large Acknowledgement}\\[1ex]

The author acknowledges fruitful discussions with Horst von Borzeszkowski. The work was supported in part by the Deutsche Forschungsgemeinschaft.


\end{document}